\documentclass[iicol, pdflatex,sn-mathphys-num]{sn-jnl}% Math and Physical Sciences Numbered Reference Style
\makeatletter

\renewcommand{\Authorfont}{%
  \reset@font
  \fontsize{12bp}{12bp}\selectfont
  \boldmath
  \centering
}

\renewcommand{\addressfont}{%
  \reset@font
  \fontsize{9bp}{11bp}\selectfont
  \titraggedcenter
}

\usepackage{etoolbox}
\makeatletter
\patchcmd{\@maketitle}
  {\artauthors\par}
  {{\centering\artauthors\par}}

\makeatother
\usepackage{graphicx}%
\usepackage{multirow}%
\usepackage{amsmath,amssymb,amsfonts}%
\usepackage{amsthm}%
\usepackage{mathrsfs}%
\usepackage[title]{appendix}%
\usepackage{xcolor}%
\usepackage{textcomp}%
\usepackage{manyfoot}%
\usepackage{booktabs}%
\usepackage{algorithm}%
\usepackage{algorithmicx}%
\usepackage{algpseudocode}%
\usepackage{listings}%
\usepackage{float}

\usepackage{etoolbox}   
\makeatletter
\patchcmd{\@maketitle}{\artauthors}{{\artauthors}}{}{}
\makeatother

\begin{document}

\title[Article Title]{A Solar Optical Clock}

\author[1,2]{\fnm{Connor} \sur{Fredrick}}\nomail
\equalcont{These authors contributed equally to this work.}

\author*[2,3]{\fnm{Ryan K.} \sur{Cole}}\email{rcole@bates.edu}
\equalcont{These authors contributed equally to this work.}

\author[4]{\fnm{Ryan} \sur{Terrien}}\nomail

\author[5,6]{\fnm{Suvrath} \sur{Mahadevan}}\nomail

\author*[1,7]{\fnm{Scott A.} \sur{Diddams}}\email{scott.diddams@colorado.edu }

\affil[1]{Electrical, Computer and Energy Engineering, University of Colorado, Boulder, CO 80309 USA}

\affil[2]{Time and Frequency Division, National Institute of Standards and Technology, Boulder, CO 80305 USA}

\affil[3]{Department of Physics and Astronomy, Bates College, Lewiston, ME 04240 USA}

\affil[4]{Department of Physics and Astronomy, Carleton College, Northfield, MN, USA}

\affil[5]{Department of Physics and Astronomy, Pennsylvania State University, University Park, PA, USA}

\affil[6]{Center for Exoplanets and Habitable Worlds, Pennsylvania State University, University Park, PA, USA}

\affil[7]{Department of Physics, University of Colorado, Boulder, CO 80309 USA}

%TC:ignore
\abstract{Precision spectroscopy underpins quantum sensing, atomic timekeeping, and the search for terrestrial-mass exoplanets. Finding an Earth-Sun analog requires fractional splitting of spectral features that rivals the precision of the best laboratory spectroscopy. This corresponds to resolution of a few parts in $10^5$ of a typical stellar spectral line, or a Doppler velocity precision of less than $10$~cm~s$^{-1}$. Only a few astrophysical spectrometers reach this regime, and only by dispersing starlight with large gratings and averaging thousands of lines. Here, we show that comparable precision can be achieved on a single solar line by leveraging techniques used in quantum sensors and optical clocks. A laser is frequency-locked to the 1565 nm solar Fe I line by coupling to a single mode of the solar thermal field and counted against an optical frequency comb referenced to the SI second, thus creating a solar optical clock.  Operating at the fundamental photon-noise limit of thermal heterodyne spectroscopy, we achieve centroid precision of one part in $10^9$ of the optical frequency, or 30~cm~s$^{-1}$ in radial velocity. We recover the Sun's five-minute acoustic oscillations and the expected Rossiter-McLaughlin signal during a solar eclipse. Combining these techniques from quantum sensing with a thirty-meter-class telescope could enable our demonstrated precision on the nearest Sun-like stars in a few hours. Multiplexing such absolute frequency measurements across lines with different magnetic sensitivity and formation depth would reduce averaging time and help disentangle stellar granulation and activity from the Doppler signature of terrestrial exoplanets, providing new approaches to address some of the most demanding measurements in astrophysical spectroscopy.
}
%TC:endignore

%\keywords{keyword1, Keyword2, Keyword3, Keyword4}



\maketitle

%%%%%%%%%%%%%%%%%%%%%%%%%%  body  %%%%%%%%%%%%%%%%%%%%%%%%%%
\section{Introduction}

\begin{figure*}[h]
\centering\includegraphics[width=\textwidth]{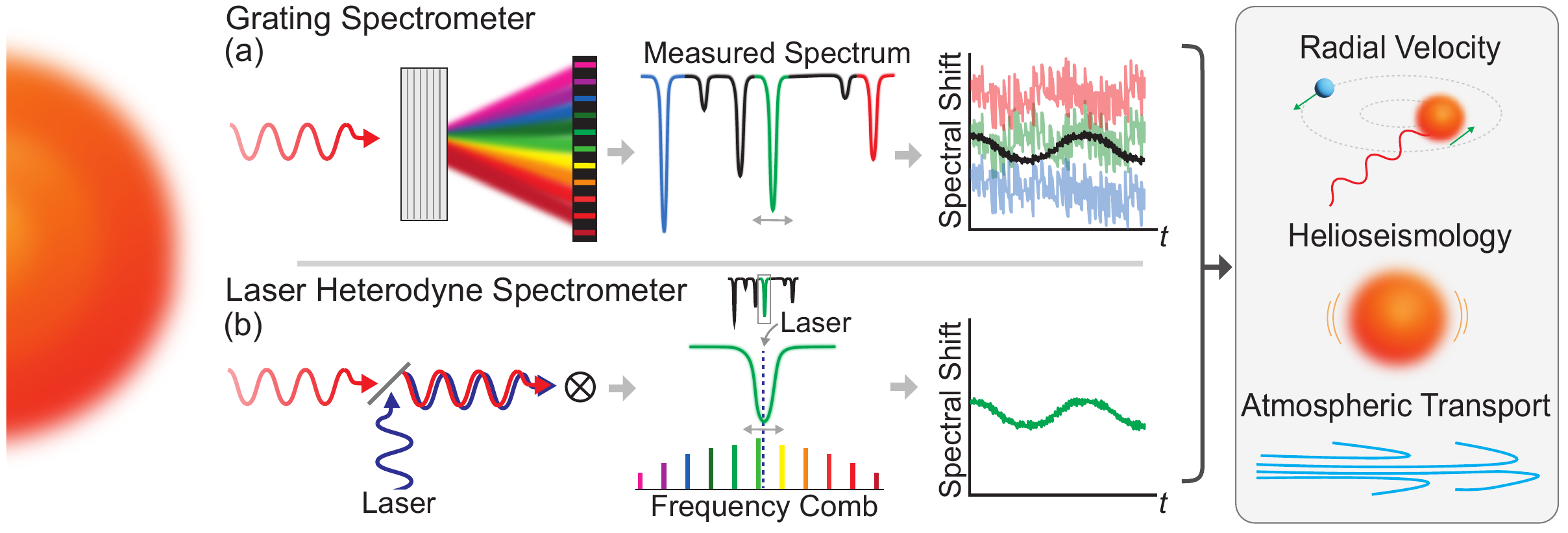}
\caption{Comparison of a grating-based astrophysical spectrometer to a laser heterodyne spectrometer designed in analogy to a modern optical clock. (a) Conceptual layout of a direct-detection grating spectrograph. The dispersed stellar spectrum is measured on a detector array, and Doppler shifts are determined by averaging the wavelength shifts of many lines across the stellar spectrum. (b)~Schematic of a solar optical clock. A probe laser is locked to the center of a transition in the solar spectrum using heterodyne spectroscopy and modulation techniques. The frequency of the probe laser is simultaneously measured relative to an optical frequency comb. The resulting frequency measurements encode Doppler velocity information that can be used to probe phenomena related to exoplanets, solar physics, and atmospheric science.}
\label{fig:concept}
\end{figure*}

Quantum sensing relies on the measurement of changes in the energy levels or state of an atom to detect external parameters such as electric or magnetic fields, time, motion, force or rotation \cite{degen2017quantum}. Many of the most precise and sensitive atomic and molecular quantum sensors rely on modern laser spectroscopy techniques of frequency modulation, precise servo control, heterodyne detection and direct counting of optical frequencies \cite{Moses:77,Bjorklund:80,Udem_2002,Haensch_Nobel,Hall_Nobel}. For example, these tools form the technological foundation of modern optical clocks, which depend on the precise synchronization of a laser oscillator to the frequency difference of quantum states in an atom \cite{ludlow2015optical, fortier2026optical}. In practice, this synchronization is realized by stabilizing a probe laser to a narrow atomic transition and using a laser frequency comb to downconvert the resulting optical frequency reference to a microwave signal that can be counted electronically.  Today, the best optical clocks reach fractional stability at the level of $10^{-18}$ or better \cite{fortier2026optical}. This makes them exquisite quantum sensors of external EM fields \cite{hassan2025cryogenic} and gravity \cite{takamoto2020test}, as well as compelling tools in searches for new physics beyond the Standard Model \cite{safronova2018search,peik2021nuclear,arakawa2026multimessenger}. 

However, these precision techniques rely on highly coherent laser sources, limiting their application to light generated in the laboratory and excluding the vast sources of thermal light from across the universe. Nonetheless, many astrophysical spectroscopic problems demand frequency precision in terms of relative linesplitting that is comparable to, or even exceeding, that of the best laboratory spectroscopy. An important example lies in the detection of subtle Doppler radial velocity (RV) signals in stellar spectra that indicate the presence of an orbiting exoplanet \cite{fischer2016state}. These measurements play a central role in efforts to understand the origins of planetary systems and in the ongoing search for Earth-like planets beyond our Solar System. 

An Earth-mass planet orbiting a Sun-like star at 1 A.U. induces a RV signal with an amplitude of only 8.9 cm/s, corresponding to fractional optical frequency shifts of a few parts in $10^{-10}$ occurring over orbital periods of a year or more \cite{fischer2016state}. Pursuing this level of sensitivity places stringent demands on instrument stability and calibration. State-of-the-art planet-hunting spectrometers have pursued this challenge with great effect \cite[e.g.][]{thompson2016harps3,schwab2016design, pepe2021espresso, jurgenson2016expres, gibson2016kpf}, generally by employing large-scale, broadband dispersive spectrographs that average hundreds or thousands of transitions in measured stellar spectra to make sensitive measurements of subtle Doppler shifts (see Figure \ref{fig:concept}). These instruments are based on the direct detection of stellar light, with Doppler shifts encoded as the position of spectral features on a detector. Although frequency combs now provide exquisite calibration for these instruments \cite[e.g.][]{wilken2012spectrograph,metcalf2019stellar}, the underlying measurement remains one of wavelength and not frequency. This distinction has historically separated these precision astrophysical measurements from analogous challenges in modern frequency metrology and quantum sensing that employ coherent laser spectroscopy, heterodyne detection and frequency counting. Leveraging these approaches, here we show that Doppler shifts can, in fact, be measured precisely as a frequency even on thermal sources such as the Sun.

\begin{figure*}[t]
\centering\includegraphics[width=\textwidth]{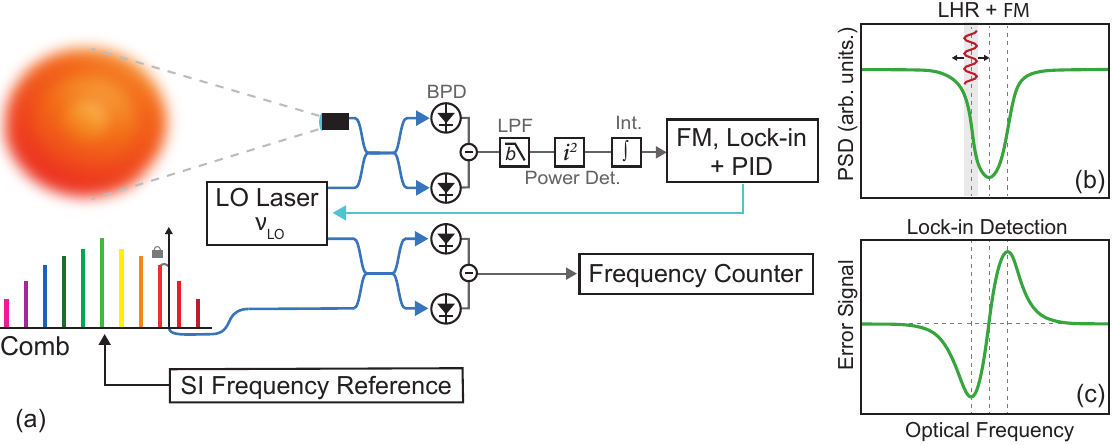}
\caption{Overview of our experimental approach. Sunlight is coupled into single-mode optical fiber and mixed with a local oscillator (LO) laser on a photodetector. The resulting heterodyne signal is measured within a defined radio-frequency bandwidth ($b$) and passed to a power detection and rectification circuit. Frequency modulation of the LO laser and lock-in detection of the heterodyne signal power generate an error signal with a zero crossing at the center of the solar line. A loop filter locks the LO to the zero crossing, and the frequency of the stabilized LO laser is simultaneously measured against an optical frequency comb.}
\label{fig:expt}
\end{figure*}

Our approach, shown conceptually in Figure \ref{fig:concept}(b), applies the power of laser and optical frequency-comb metrology to astrophysical spectroscopy and is inspired by the architecture of modern optical clocks and other quantum sensors. However, instead of carefully isolating an atom from its environment, we exploit the interactions between atoms and their environment to probe the magnetic and dynamic activity of the Sun. Specifically, we use thermal heterodyne and frequency modulation and demodulation to lock the frequency of a laboratory laser to the center of a Fe~I solar transition near 1565.28~nm. Once locked, the laser frequency is counted with an optical frequency comb that is referenced to the SI second, thereby realizing the main components of what could be considered a solar optical clock. Our technique builds on earlier work of wavelength-modulated and comb-calibrated laser heterodyne radiometry (LHR) \cite{fredrick2022thermal,Martin-Mateos:18, cole2025frequency}, and allows us to improve the line splitting of the 4.3~GHz linewidth of the Fe~I transition to 1 part in 1,400 in one second of measurement. With averaging, the frequency precision improves to 200~kHz, the equivalent of 30~cm/s in Doppler radial velocity or one part in $10^{9}$ of the absolute optical frequency. The measured sensitivity reaches the quantum-noise limit expected for thermal heterodyne detection, providing an experimental verification of the precision limits for optical-heterodyne-based radial velocity measurements on a single line.

Our combination of modern laser spectroscopy techniques with traditional heterodyne radiometry \cite{parvitte2004infrared, menzies1977ozone, kostiuk1983remote, nieuwenhuijzen1970optical} opens new possibilities for precision astrophysical spectroscopy and sensing. As an example, we study the change in the rotational Doppler shift of the Sun during a partial solar eclipse in 2024, observing the Doppler signature of the Rossiter-McLaughlin effect \cite{rossiter1924detection,mclaughlin1924some} with meter-per-second precision.  With its excellent frequency precision and high signal-to-noise ratio (SNR), our technique is suited for other solar observations that would be accessible at high spatial resolution--such as probing convective motions, eruptive events, sunspots, the solar magnetosphere and wind. Such solar observations also provide a bridge for interpreting stellar spectra \cite{zhao2023extreme}. A combination of Sun-as-a-star (e.g., solar disk integrated) and spatially resolved observations could also help understand the underlying causes of observed spectral variations that are one of the most challenging problems in exoplanet science \cite{fischer2016state, zhao2022expres, ellwarth2023iag}. Importantly, this approach is not intrinsically limited to studies of the Sun. We show that sensitivity at the quantum limit is feasible on stars beyond the Sun. This could be accomplished with a 25--40~m class, adaptive-optics-fed telescope, making sensitive, single-line RV measurements possible on the nearest Sun-like stars. Beyond astronomy, the same setup can be used for sensing and velocimetry in applications ranging from atmospheric monitoring \cite{Cole:23} to spaceflight.

\section{Experimental Setup}

Figure \ref{fig:expt}(a) illustrates the main components of the experiment, and further details are found in Appendix \ref{apA}. Sunlight is coupled into single-mode fiber (SMF) using an integrated fiber collimating package that is mounted on an automated solar tracker. A refractive beam-shaping optic placed in front of the fiber collimator is used to generate a flat-top antenna pattern in the far field to uniformly integrate light from across the solar disk. A complete description of the solar tracking and fiber coupling design is given in Refs. \cite{fredrick2022thermal,Cole:23, cole2025frequency}. Sunlight coupled into the fiber has blackbody limited power spectral density on the order of $3\times10^{-20}$ W/Hz in a single polarization, which is then heterodyned with a local oscillator (LO) laser on a balanced photodetector. 

The downconverted power from within the optical bandwidth $\Delta \nu=2b$ (set by a radio-frequency low pass filter with bandwidth $b=1.65$~GHz) is integrated and sent to a lock-in detection circuit. The LO laser frequency is modulated with a 5~kHz sine wave having peak-to-peak modulation of $2\Delta f=4$~GHz. The detected LHR signal is demodulated at the same frequency for 1st-order lock-in detection. As shown in Fig.~\ref{fig:expt}(b)-(c), when the center frequency of the modulated laser is scanned across the region of iron absorption, the demodulated quadrature phase results in the derivative-like error signal that is used to lock the LO laser to the center of the solar transition.  %An electro-optic intensity modulator stabilizes the LO power as the laser current is dithered.

Once locked, the LO optical frequency is counted with the aid of an optical frequency comb that has its offset frequency and 250~MHz repetition rate stabilized to a frequency reference traceable to the SI second \cite{jones2000carrier}. To eliminate ambiguity at integer multiples of 250~MHz and to increase the spectral range of the heterodyne, the LO frequency was read out with the help of an auxiliary reference laser locked to a specified mode of the frequency comb \cite{cole2025frequency}. The LO and reference lasers were then heterodyned on a 5~GHz balanced detector. An offset of 2.5~GHz between the two lasers ensured that the swing of the frequency modulation was within the bandwidth of both the balanced detector and the zero deadtime counter that was employed to continuously count the laser frequency.

\begin{figure}[t]
\centering\includegraphics[width=1\columnwidth]{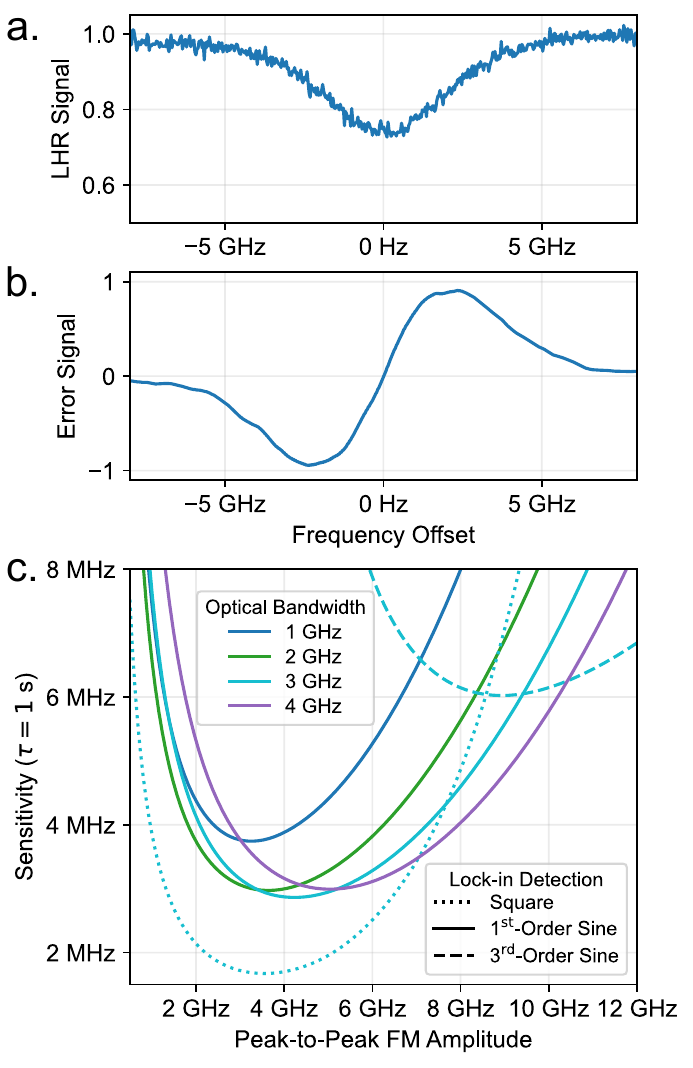}
\caption{(a) Sweeping the LO frequency reveals the solar iron line at 1565.28 nm. (b) The 1st-order error signal with a zero crossing at the center of the line. (c) Theoretical sensitivity vs. FM dither amplitude, assuming a total quantum efficiency of $\eta_{tot}\approx0.7$ and a mean photon occupancy of $\langle n \rangle \approx 0.25$ photons/s-Hz. The solid lines show the change in sensitivity for 1st-order lock-in detection with different integrated optical bandwidths $\Delta\nu$, and the dashed and dotted lines show the change in sensitivity with 3rd-order lock-in detection and square-wave lock-in detection respectively.}
\label{fig:sensitivity}
\end{figure}

Figure \ref{fig:sensitivity}(a)-(b) shows typical results that characterize the apparatus. Sweeping the LO frequency reveals the solar iron line at 1565.28~nm measured relative to the absolute and stable frequency axis provided by the frequency comb. The iron transition has an absorption depth of approximately 28\% and full width at half maximum of $\sim$4.3~GHz. As described above, demodulation provides an error signal with a zero crossing at the center of the line for locking. The projected sensitivity in determining the solar iron line center in 1~s of averaging depends on the choice of FM dither amplitude, integration bandwidth, and lock-in detection technique. Based on the experimentally accessible modulation parameters and the measured line shape, we numerically derive the sensitivity curves shown in Figure \ref{fig:sensitivity}(c). Here we see that the best (lowest) sensitivity with sinusoidal modulation is approximately 3~MHz in 1~s of averaging. This would correspond to splitting the line by approximately 1 part in 1,400, and is achieved with parameters similar to those used in our experiment ($2b=3.3$ GHz and $2\Delta f=4$ GHz). These projections match results of our experiments and are consistent with analytic expressions of shot-noise-limited frequency precision (derived in Appendix \ref{apB}). Figure \ref{fig:sensitivity}(c) also shows that the most sensitive detection technique would involve square-wave modulation, where the laser only samples the lineshape at the extrema of the derivative that maximizes the frequency discrimination. This is the same modulation scheme used in the best optical clocks \cite{ludlow2015optical}.

\section{Results}

\subsection{Precision Radial Velocity Measurements of the 1565 nm Solar Fe I Transition}

\begin{figure}[t]
\centering\includegraphics[width=1\columnwidth]{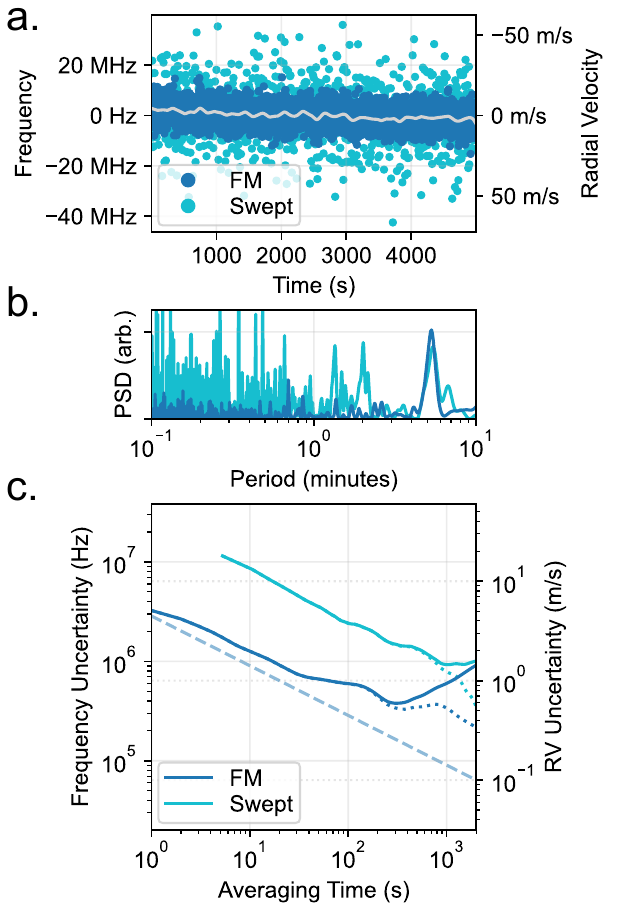}
\caption{Comparison of FM-locked and swept LHR. (a)~Frequency shifts observed over 5000 seconds of continuous measurements. The SNR-efficient FM LHR (dark blue) reaches a higher precision in a shorter amount of time than swept LHR (light blue). A small drift is present due to coupling between the apparent orientation of the Sun’s rotational axis and a slight asymmetry in the antenna pattern of the light collection optics. Binning of the FM data to 100 s intervals yields the smoothed grey line. (b)~The power spectral density (PSD) of these binned data clearly reveals the 5.5 min oscillation period characteristic of solar p-modes. (c)~The Allan deviations of the FM and swept LHR data. Again, evidence of solar p-modes is present for the FM data, which shows a peak in the averaging time at half the p-mode period ($\sim$ 150 s). The dotted lines represent the Allan deviation after linearly detrending the frequency shifts, and the dashed line indicates the theoretical quantum-limit for FM-locked LHR sensitivity discussed in Section 3 and derived analytically in the Appendix.}
\label{fig:precision}
\end{figure}

Stellar radial velocity measurements provide an ideal test case for applying precision frequency metrology techniques to astrophysical observations. In this context, we use our measurements of the 1565.28~nm solar iron line to track the radial velocity of the Sun and to probe the complex dynamics of the solar atmosphere. Figure~\ref{fig:precision}(a) shows these Doppler shift measurements over a representative day, and Figure 4(b) shows the power spectral density (PSD) of the measured shifts after binning to 100 second intervals. The PSD reveals a significant peak at the 5.5 minute period characteristic of solar acoustic oscillations (p-modes). These oscillations are a well known probe of the interior structure and dynamics of the Sun \cite{chaplin2007solar}, and have been studied extensively by long running Doppler velocity observatories such as BiSON \cite{broomhall2009definitive}, GONG \cite{jain2026three}, and the Helioseismic and Magnetic Imager aboard the Solar Dynamics Observatory \cite{scherrer2012helioseismic}. They are also an important source of variability in both solar and stellar radial velocity measurements \cite{crass2021extreme}. 

The Allan deviation of the measured frequency shifts is shown in Figure \ref{fig:precision}(c). Our approach tracks the center of the 1565~nm transition with a precision of $\sim$3~MHz (4.7 m/s) in one second of averaging, equivalent to splitting the 4.3~GHz linewidth by a factor of 1,400. This experimental sensitivity is in excellent agreement with the numerical simulations of Figure \ref{fig:sensitivity} and the analytic predictions discussed below and in the Appendix. The measured frequency shifts exhibit a weak linear trend that we attribute to coupling between the apparent orientation of the Sun’s rotational axis and a slight asymmetry in the antenna pattern of the light collection optics (see Ref. \cite{cole2025frequency}). After linearly detrending to correct for this drift, our measurements support a precision approaching 200~kHz ($\sim$30 cm/s in radial velocity) after a few thousand seconds of averaging, equivalent to one part in 10$^9$ of the absolute optical frequency and 5 parts in $10^5$ of the linewidth. The influence of p-mode oscillations is also evident in this view of our measurements, which shows a clear increase in the frequency precision after averaging over the five-minute (300~s) oscillation period.%, due to the drop in noise when averaging the p-modes.

To place these results in context, we compare our frequency-locked measurements to simultaneous measurements acquired with a more conventional swept laser heterodyne radiometer that fully resolves the lineshape of the same Fe~I transition \cite{fredrick2022thermal, cole2025frequency}. The swept LHR instrument uses the same solar tracking and fiber coupling hardware as the frequency modulated approach described above, and both instruments record simultaneously by splitting the fiber-coupled sunlight in a fiberized polarizing beam splitter. The principal difference between the two instruments lies in their measurement strategy. The frequency-locked approach continuously tracks the line center, while the swept instrument slowly scans across the solar transition in order to precisely resolve the spectral profile. Doppler shifts are determined from the swept measurements in post processing as the frequency offsets that maximize the cross-correlation between individual spectra and a template.  These shift measurements and their Allan deviation are shown in Figure \ref{fig:precision}.

To quantify the frequency uncertainty shown in Figure \ref{fig:precision}, it is useful to consider how the precision of a line center measurement scales with the properties of the observed spectral feature. For both approaches, the precision of each frequency measurement scales as \cite{ludlow2015optical}
\begin{equation}
\label{eq:scaling}
\sigma_\nu \approx \frac{\delta\nu \times\beta }{D\times {\rm SNR}_\tau}
\end{equation}
where $\sigma_\nu$ is the uncertainty in the line center frequency at time $\tau$, $\delta\nu$ is the linewidth of the measured transition, $D$ is the line depth, and $\beta$ is a numerical factor accounting for the shape of the line and how it is sampled by the instrument. The quantity SNR$_\tau$ is the shot-noise-limited signal-to-noise ratio obtained over the same averaging time $\tau$, and is derived rigorously for thermal heterodyne spectroscopy in Ref. \cite{zmuidzinas2003thermal}. 

When used with the experimental parameters of the FM-locked measurements, Eq. \ref{eq:scaling} estimates a one-second frequency precision of 3~MHz. This is in excellent agreement with the measured sensitivity shown in Figure \ref{fig:precision} and the numerical estimate of Figure \ref{fig:sensitivity}.  The agreement highlights that the sensitivity of our measurements is limited only by the photon statistics of the observed thermal field. As such, these results can be viewed as an experimental test of the precision limits for optical-heterodyne-based radial velocity measurements on a single line. Any further improvement would require changes to the measurement strategy, such as adopting a more information-efficient modulation approach (e.g., the square-wave modulation shown in Figure \ref{fig:sensitivity}). %The same framework can also be extended to consider future observations of stellar targets other than the Sun, discussed below. 

Equation \ref{eq:scaling} also provides a basis to understand the sensitivity difference between the FM-locked and swept LHR measurements. The most significant differences between the two approaches arise from their heterodyne SNR and their allocation of measurement time across the target transition. In this context, the FM method is advantageous because it both increases the heterodyne SNR through a larger integration bandwidth and also concentrates the measurement on the steepest part of the spectral line (where sensitivity to frequency shifts is greatest). In contrast, the swept LHR approach uses a lower integration bandwidth in order to precisely resolve the line shape while also distributing the measurement over a wider range of optical frequencies such that only a portion of each scan samples the region of the spectrum that conveys frequency shift information. Taken together, these factors lead to a substantial improvement in frequency precision for the FM-locked approach that is consistent with the near-order-of-magnitude difference shown in Figure \ref{fig:precision} (see further discussion in Appendix \ref{apB}).

\subsection{Rossiter-McLaughlin Observations of the 2024 Solar Eclipse}

\begin{figure*}[t]
\centering\includegraphics[width=1\textwidth]{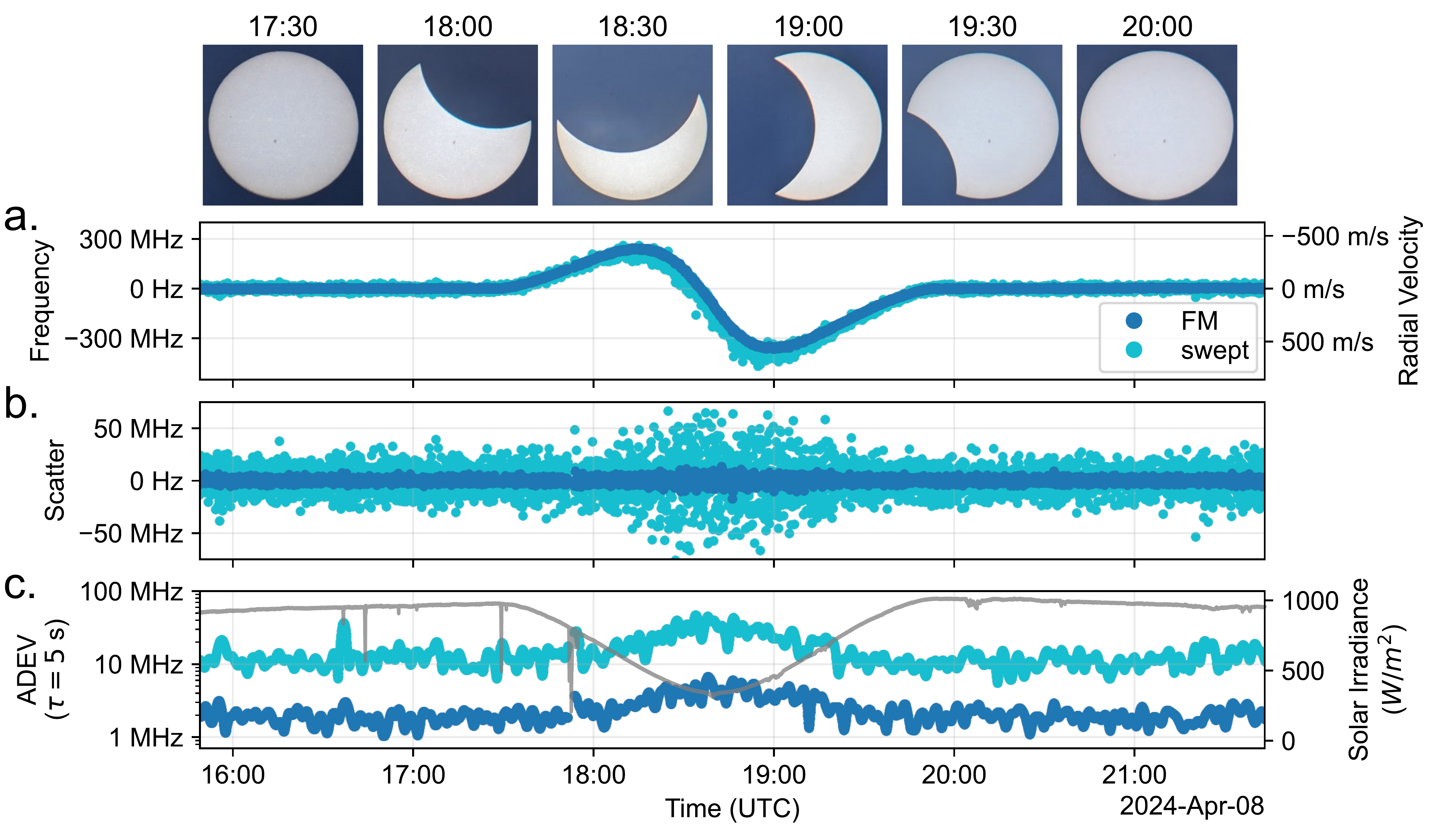}
\caption{Simultaneous FM-locked and swept LHR data taken during the 2024 solar eclipse in Boulder CO, USA. (a) Optical frequency shifts measured during the eclipse, which clearly show the expected Rossiter-McLaughlin signature produced by the partial occultation of the solar disk. (b) Detrended frequency shifts between the swept and FM-locked measurements, illustrating the relative scatter between the two approaches. (c) Five-second Allan deviation for the FM-locked and swept measurements plotted as a function of time during the eclipse. Also shown is the independently measured solar irradiance which decreases during the eclipse, thereby increasing the frequency scatter and Allan deviation.}
\label{fig:eclipse}
\end{figure*}

As a further demonstration of our approach, we used both the FM-locked and swept LHR instruments to observe the April 2024 partial eclipse in Boulder, CO, USA, where peak obscuration reached $\sim$80\%. Solar Doppler spectroscopy during an eclipse presents a unique opportunity to explore both the complexity of precision astrophysical spectroscopy and the information that can be recovered from these measurements. Under normal observing conditions of Sec. 3.1, our measurements depend on the rotational symmetry of the Sun to cancel rotational Doppler shifts from opposing limbs, thereby enabling sensitivity to center-of-mass Doppler shifts and signatures of variability (e.g. p-modes) integrated across the solar disk. An eclipse breaks this symmetry by progressively occluding light from different regions of the solar disk. The resulting imbalance introduces a time-dependent Doppler shift known as the Rossiter-McLaughlin effect \cite{rossiter1924detection,mclaughlin1924some} that is used in studies of exoplanetary systems \cite{gaudi2007prospects}, the Sun \cite{reiners2016radial}, and moons in the Solar System \cite{molaro2013detection}.

Figure \ref{fig:eclipse}(a) shows the frequency shifts in the 1565~nm Fe I transition during the April 2024 eclipse recorded with both the frequency-locked and swept LHR. A linear trend has been removed from both datasets to account for the pointing-induced drift discussed above and in Ref. \cite{cole2025frequency}. The measured frequency shifts clearly show the expected Rossiter-McLaughlin signal \cite{reiners2016radial, gonzalez2026inprep} with an amplitude of $\sim$500~m/s. Panel (b) shows the measured radial velocities after detrending with a smoothing function, which highlights the improved precision achieved by continuously tracking the line-center frequency. This improvement is also shown in Panel (c), which shows the 5~s Allan deviation of the frequency shift measurements as a function of time during the eclipse. Under normal conditions, the FM-locked approach achieves nearly an order-of-magnitude improvement in precision relative to the swept measurements. During the eclipse, the incident solar flux decreases as the Moon progressively obscures the solar disk. This reduces the heterodyne SNR and produces a corresponding increase in the Allan deviation that closely tracks the changing solar irradiance measured by a pyrheliometer mounted to the solar tracker. 

The shape of the Rossiter-McLaughlin curve encodes information about the size of the transiting body (the Moon) and the relative angle between its path and the spin axis of the star (the Sun), and can also provide details about solar surface inhomogeneities and convective dynamics. While we do not attempt to extract this information here, our results demonstrate that our combination of coherent laser spectroscopy techniques with traditional heterodyne radiometry can measure these signals with sub-meter-per-second sensitivity on a single line. Such observations should offer a powerful tool for detailed studies of the Sun and its atmosphere, and for connecting these studies to the unresolved stellar observations that underpin modern radial velocity surveys.

\section{Scaling to Other Astrophysical Targets}

\begin{figure*}[t]
\centering\includegraphics[width=0.7\textwidth]{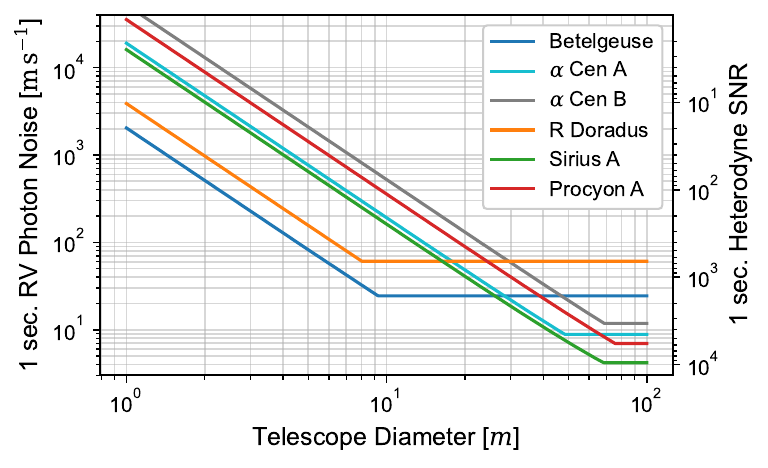}
\caption{Estimated 1~s radial-velocity precision for frequency-locked LHR measurements of selected large angular diameter stars as a function of telescope diameter. For ease of comparison, we assume the same linewidth and absorption depth as the 1565~nm solar Fe I line studied in this work. Diffraction-limited 25-40~m class telescopes could enable radial-velocity precision on a single line at the level of tens m/s or better for a number of stellar targets.}
\label{fig:otherstars}
\end{figure*}

By establishing the fundamental sensitivity limits for optical heterodyne based radial velocity measurements on the Sun, our same framework can be extended to consider observations of other astrophysical targets. The achievable RV precision for our FM LHR approach (Equation \ref{eq:scaling}) is driven by the properties of the target transition (linewidth and line depth) and by the heterodyne SNR. We can estimate the heterodyne SNR for observations of other stellar targets with large, adaptive-optics (AO) fed telescopes by expanding the total optical efficiency term in the heterodyne SNR (Equation \ref{eq:SNR}) as 
\begin{equation}
    \label{eq:eta}
\eta_{tot}=\eta_\Omega\eta_{SM}\eta_{opt}
\end{equation}
where $\eta_{opt}$ accounts for the optical efficiency of the instrument, $\eta_{SM}$ accounts for the coupling between the telescope and the single-mode heterodyne receiver, and $\eta_{\Omega}$ describes the angular filling factor of the source within the receiver antenna pattern. 

Coupling efficiencies between large, AO-fed telescopes and single-mode fibers have been experimentally demonstrated at the level of 0.5 near 1550~nm \cite{jovanovic2017efficient}. The angular filling factor $\eta_{\Omega}$ is defined as $\Omega_S/\Omega_A$ (for $\Omega_S<\Omega_A$) and unity for $\Omega_S\geq\Omega_A$ \cite{blaney1975signal}, where $\Omega_S$ and $\Omega_A$ are the solid angles subtended by the source and antenna pattern, respectively. The antenna theorem constrains the product of the antenna solid angle and the effective telescope aperture ($A$) as $A\times\Omega_A=\lambda^2$ \cite{siegman1966antenna}. With this, $\eta_{\Omega}$ can be expressed approximately in terms of the angular diameter of the source ($\Theta$) and the telescope diameter (D)
\begin{equation}
\label{eq:etaomega}
\eta_{\Omega}\approx \left(\frac{\pi D\Theta}{4\lambda}\right) ^2
\end{equation}
which is valid in the approximate limit that the source intensity profile and the receiver antenna pattern are uniform within the field of view. 

Combining Equations \ref{eq:scaling}--\ref{eq:etaomega} and \ref{eq:SNR} allows for an estimate of the achievable RV sensitivity on other stellar targets. These predictions are shown as a function of telescope diameter in Figure \ref{fig:otherstars} for a subset of large angular size giant and main-sequence stars. These estimates assume stellar temperatures based on spectral type and angular diameters measured with interferometry \cite[e.g.][]{kervella2017_1}. For simplicity, the calculations also assume a common target transition with properties (linewidth and absorption depth) similar to the 1565~nm solar Fe I transition studied in this work. 

Figure \ref{fig:otherstars} indicates that FM LHR measurements could be feasible for a range of stellar targets when paired with a large, AO-fed telescope. In particular, 25--40~m class telescopes, a range bounded by the Giant Magellan Telescope and the Extremely Large Telescope currently under construction, enable single-line RV measurements with sensitivity at the level of tens of m/s or better in one second of averaging (i.e., comparable to the 1 second sensitivity demonstrated on the Sun in this work). Square-wave modulation (Fig.~\ref{fig:sensitivity}), long term averaging, and multiplexing over multiple spectral lines, would bring accurate single-line stellar RV measurements at the 10 cm/s level into the realm of possibility. On real stars, however, granulation imposes an astrophysical floor of several tens of cm/s, even on hour-long timescales \cite{zhao2023extreme}. Highly precise and accurate line-by-line measurements are needed to decouple the complex changes in magnetic activity and convective velocity fields from the Doppler shift due to an orbiting planet. Recent work has shown the promise of using sets of lines to distinguish these stellar effects from the signature of an exoplanet \cite{dumusque2018measuring, al2024measuring}. By providing long-term high-accuracy frequency measurements traceable to the SI second itself, observations with the method presented here enable the possibility of disentangling stellar variability and exoplanet-induced Doppler signals at the individual line level for the highest-value stellar targets.

Beyond exoplanet detection, the next generation of high-precision spectrometers on large aperture telescopes will also inform questions of fundamental physics, including the first direct measurements of the expansion history of the universe \cite{quercellini2012real}, searches for time variation of fundamental constants \cite[e.g.][]{trost2025espresso}, and studies of the local distribution of dark matter \cite{ravi2019probing}. These efforts hinge on measurements of individual line centroids and shifts that remain accurate at the $\sim$1-10~cm/s level over observing baselines that may extend for a decade or longer. Large aperture telescopes will open the door to new targets for these measurements, but exploiting that capability will require limiting instrumental systematics to the level of the photon noise as well as maintaining accurate frequency scales across instruments and observing epochs, both challenging tasks for even the most stable, comb-equipped dispersive spectrometers (e.g. ANDES \cite{martins2024cosmology}). The current generation of precision radial velocity spectrometers has already shown the value of using contemporaneous observations of bright stars and the Sun for establishing accuracy and for cross-calibration of measurements across instruments and epochs \cite{zhao2023extreme}. In this sense, by enabling high-accuracy single-line spectroscopy of these targets, the method we present here could provide a valuable and independent part of the cross-calibration portfolio for next-generation spectrometers with a direct tie to the SI second.

\section{Conclusions}
In this paper, we have shown how to translate the powerful techniques of FM spectroscopy and laser frequency locking and counting from the field of optical atomic clocks to precision astrophysical spectroscopy. These techniques immediately yield nearly an order of magnitude improvement in the frequency precision of laser heterodyne measurements of Doppler radial velocities, with still further gains possible. In these first applications, this allows us to determine and track the line center of a single solar absorption feature with quantum-limited SNR and frequency precision of 1 part-per-billion, or equivalently 30~cm/s in Doppler radial velocity. By enabling direct frequency measurements, the relatively simple instrumentation we introduce achieves comparable precision on a single spectral line to what highly stabilized and expensive state-of-the-art astronomical spectrographs can achieve after aggregating Doppler information from thousands of lines. This capability opens new prospects for using combinations of individual lines to detect and remove stellar activity noise in radial velocity measurements. 

More broadly, our combination of modern laser spectroscopic techniques with traditional heterodyne radiometry should provide new opportunities for precision astrophysical spectroscopy and sensing. While we illustrate these advantages in solar spectroscopy, these approaches could provide a bridge between solar precision Doppler spectroscopy and a broader range of astrophysical measurements.  The improvements in sensitivity we introduce could be of particular interest when combined with advances in infrared astronomy and adaptive optics on 10--40-meter class telescopes that would be required for single-mode imaging of nearby stars and larger astrophysical objects with sufficient signal-to-noise ratio. Such facilities could extend frequency-locked LHR to sensitive, single-line radial velocity measurements of favorable stellar targets, with further improvement possible through longer averaging and multiplexing across multiple spectral lines. While many challenges remain, tightening the connection between the communities of laser-based laboratory metrology and astrophysical precision spectroscopy promises to hold future advances for both.

\backmatter

%TC:ignore
\bmhead{Funding}
This work is supported by NIST Innovations in Measurement Science program, NASA grants  80NSSC21K2006 and 80NSSC26K0554, the W.M. Keck Foundation, and the Air Force Office of Scientific Research under the PRISM MURI award number FA9550-24-1-0349. R.K.C. acknowledges support from the NRC postdoctoral fellowship.

\bmhead{Author Contributions}
C.F. and R.K.C. built the experiment and collected the data. S.A.D. conceived and supervised the project. C.F. analyzed the data. C.F. and R.K.C. interpreted the data and developed the Doppler sensitivity framework. R.T. and S.M. provided astrophysical interpretation of the data and insight into other applications. R.K.C developed the scaling relations for other astrophysical targets. R.K.C., C.F., and S.A.D. wrote the manuscript. R.K.C. and C.F. developed the figure set. All authors contributed to the editing of the paper and discussion of the results.

%\bmhead{Data Availability}
%The data supporting the figures in this paper will be made available at \url{10.5281/zenodo.23165213} upon publication. 
%TC:endignore

%TC:ignore
\begin{appendices}
\section{Experimental Methods}\label{apA}

\begin{figure*}[t]
\centering\includegraphics[width=0.9\textwidth]{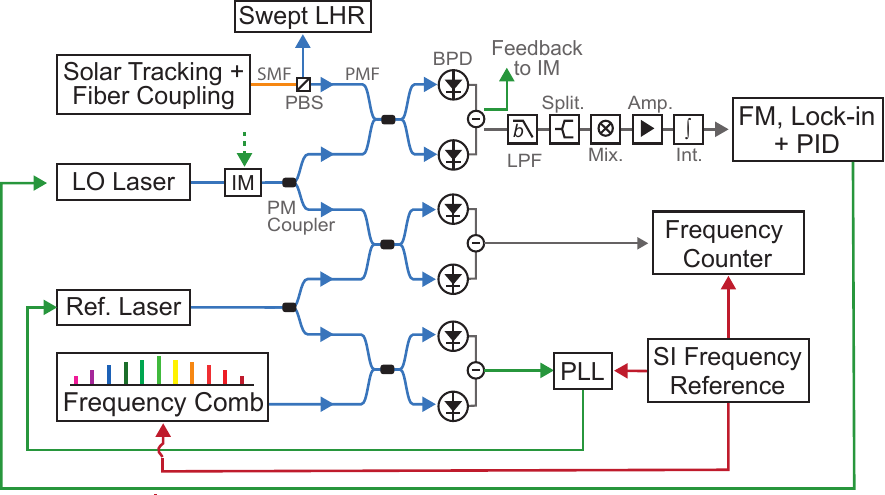}
\caption{Detailed experimental schematic of the frequency modulated LHR system. With the exception of the solar tracking and fiber coupling optics, all components are located in a rooftop laboratory $\sim$20~m from the solar tracker. SMF: single-mode optical fiber, PMF: polarization-maintaining optical fiber, BPD: balanced photodetector, IM: electro-optic intensity modulator, LPF: low-pass filter, Amp: RF amplifier, Split: RF power splitter, Mix: RF mixer, Int: integrator, PLL: phase-lock loop.}
\label{fig:expt_methods}
\end{figure*}

Figure \ref{fig:expt_methods} shows a detailed schematic of our experimental apparatus. Light from the Sun is coupled into single-mode fiber (SMF-28) using the solar tracking and fiber coupling design described in Refs. \cite{cole2025frequency, Cole:23, fredrick2022thermal}. The fiber-coupled solar light travels through $\sim$20~m of single-mode fiber to a rooftop laboratory where it is split in a fiberized polarizing beam splitter. One linearly polarized output of the beam splitter is directed to the swept LHR apparatus \cite{cole2025frequency, Cole:23, fredrick2022thermal}, while the other is combined with light from the local oscillator laser in a polarization-maintaining 50:50 coupler and detected on a balanced photodetector (ThorLabs PDB780CAC, 2.5~GHz bandwidth). The local oscillator laser is a distributed feedback laser manufactured by Eblana Photonics with a linewidth of $\sim$3~MHz. 

The DC output of the balanced detector is used to feed back to an electro-optic intensity modulator that stabilizes the LO optical power throughout the frequency modulation and sweeping cycle. The radio frequency (RF) output is passed to the heterodyne power detection circuit described in Ref. \cite{cole2025frequency,fredrick2022thermal}. Briefly, the measured heterodyne signal is filtered to a bandwidth $b=1.65$~GHz, which also sets the optical integration bandwidth of the receiver to $2b$. The filtered RF power is detected using the combination of a 0$^{\circ}$ power splitter and a double-balanced mixer. The mixer output is amplified and integrated with a 20~kHz low pass filter before being passed to a lock-in detection module. Frequency modulation, lock-in detection, and servo control of the LO laser are implemented on a Red Pitaya Field Programmable Gate Array (FPGA) platform using the application developed by Ref. \cite{luda2019compact}. The LO laser frequency is modulated with a 5~kHz sine wave with peak-to-peak modulation of $\sim$4~GHz. 

Once locked to the target transition, the LO is heterodyned with a second CW reference laser that is phase locked to a known mode of an optical frequency comb ($f_{r} = 250 \textrm{ MHz}$). The repetition rate and carrier-envelope offset frequency of the frequency comb itself are referenced to absolute frequency standards. The two lasers are combined on a second 5~GHz balanced photodetector (ThorLabs BDX3BA), and the resulting beat frequency is counted with a zero-dead-time frequency counter (Keysight 53230A). As discussed in the main text, this two-step frequency measurement avoids ambiguity associated with modulating the LO laser over a frequency range that exceeds the repetition rate of the comb. An additional advantage is that it permits measurements of Doppler shifts substantially larger than the comb mode spacing, including the Rossiter-McLaughlin shifts reported in Fig. \ref{fig:eclipse} and the barycentric velocity shifts associated with the motion of the observatory relative to the Sun (the latter is removed in post processing \cite{wright2020barycentric,kanodia2018python}). All frequency measurements, including the optical frequency comb, phase-lock loops, and frequency counter, are referenced to a NIST-calibrated hydrogen maser---thereby providing frequency traceability to the SI second.

\section{Doppler Sensitivity}\label{apB}

To estimate the Doppler sensitivity of our approach, we relate the noise in the lock error signal to the uncertainty in the measured line-center frequency. We assume the instrument measures a spectrum of the form $T(x)=1-Df(x)$ where $T$ is the transmittance, $D$ is the depth of the transition and $f(x)$ is the normalized line shape, and we define $x=(\nu-\nu_0)/\delta\nu$, where $\nu$ is the LO laser frequency, $\nu_0$ is the line center frequency and $\delta\nu$ is the linewidth \cite{silver1992frequency, schilt2003wavelength}. After modulation and lock-in detection at frequency $\omega_m$, the error signal $e(x)$ is given by
\begin{equation}
e(x) = G\left \langle T\left (x+\frac{\Delta f}{\delta\nu}cos(\omega_m t)\right )\times cos(\omega_mt) \right \rangle
\end{equation}
where $\Delta f$ is the modulation amplitude, $G$ is the lock-in gain, and the average is taken over a modulation cycle. This result is consistent with the theory of modulation spectroscopy presented in Ref. \cite{supplee1994theoretical}. 

For measurements of the line center frequency $\nu_0$, the precision of each measurement is given by 
\begin{equation}
    \label{eq:noiseslope}
\sigma_{\nu_0}  = \frac{\sigma_e}{\frac{\partial e}{\partial\nu_0}\big{|}_{\nu_0}}
\end{equation}
where $\sigma_e$ is the noise in the measured error signal, which we note is related to the heterodyne signal-to-noise ratio (discussed below) as $\sigma_e = G/SNR_{het}$. Evaluating the derivative yields the result shown in Equation \ref{eq:scaling}, with 
\begin{equation}
    \label{eq:beta}
\beta = \left \langle f'\left(\frac{\Delta f}{\delta\nu }cos(\omega_mt)\right )\times cos(\omega_mt) \right \rangle^{-1}
\end{equation}
where $f'$ denotes a derivative with respect to $x$. 

Using Equation \ref{eq:scaling} together with the measured line shape, we estimate a Doppler sensitivity of $3~{\rm MHz}/\sqrt{\tau}$ for our approach, which is in agreement with both the measured sensitivity shown in Fig. \ref{fig:precision} and the numerical estimates of Fig. \ref{fig:sensitivity}. These calculations assume a peak-to-peak modulation amplitude of 4~GHz and a modulation frequency of 5~kHz. The linewidth and depth are determined from the high-resolution swept LHR measurements and convolved with the 3.3~GHz optical integration bandwidth. 

The heterodyne SNR is given by 
\begin{equation}
{\rm SNR_{het}} = \frac{\eta_{tot} \langle{n}\rangle}{1+\eta_{tot} \langle{n}\rangle}\sqrt{\Delta\nu \tau}
\label{eq:SNR}
\end{equation}
where $\eta_{tot}$ is the overall optical efficiency, $\langle{n}\rangle$ is the mean photon occupancy of the thermal source (given by the Bose-Einstein distribution), $\Delta \nu$ is the optical integration bandwidth of the heterodyne signal, and $\tau$ is the integration time. This expression is derived rigorously in Ref. \cite{zmuidzinas2003thermal} by modeling the heterodyne receiver as a phase-insensitive, quantum-limited amplifier. Our SNR calculations assume an efficiency $\eta_{tot}\approx0.7$ and a solar disk temperature of 5772~K. %The close agreement between the measured and predicted sensitivities indicates that our frequency-locked approach operates at the fundamental limit imposed by the photon statistics of the thermal light.

This framework also provides insight into the sensitivity difference between the two LHR approaches. The frequency-locked measurements employ a significantly larger integration bandwidth when compared to the swept measurements (3.3~GHz versus 230~MHz). In the context of Equation \ref{eq:scaling}, the larger integration bandwidth adversely affects the line width and depth, but substantially increases the heterodyne SNR. Furthermore, the swept measurements distribute their measurement time across a frequency range ($\sim$34 GHz) that is substantially broader than the linewidth of the transition ($\sim$4.3 GHz). For line-center measurements, the greatest sensitivity resides near the steepest portions of the line. Consequently, only a fraction of each swept measurement samples the region of the spectrum that conveys frequency shift information. This effectively reduces the averaging time available for frequency shift measurements and introduces an additional sensitivity penalty that is absent from the frequency-locked measurements (which, by definition, continuously sample the region near line center). If estimated numerically, the combination of these effects implies a sensitivity difference by a factor of $\sim9$, which is in good agreement with the difference observed in Figure \ref{fig:precision}.

\end{appendices}
%TC:endignore

%TC:ignore
%%%%%%%%%%%%%%%%%%%%%%% References %%%%%%%%%%%%%%%%%%%%%%%%%
\bibliography{references}
%TC:endignore

\end{document}